\documentclass[3p,times,twocolumn]{elsarticle}
\usepackage{ecrc}
\volume{00}
\firstpage{1}
\journalname{Submitted to Nuclear and Particle Physics Proceedings}

\runauth{R. L. Pearson and C. Voisey}
\jid{nppp}
\jnltitlelogo{Nuclear and Particle Physics Proceedings}
\usepackage{amssymb}
\usepackage[figuresright]{rotating}
\newcommand{\cov}{\mathrm{cov}}
\newcommand{\corr}{\mathrm{corr}}
\usepackage{amsmath} 

\begin{document}

\begin{frontmatter}

\dochead{}

\title{Towards parton distribution functions with theoretical uncertainties$^*$}

\cortext[cor0]{Talk given at the 21st International Conference in Quantum Chromodynamics (QCD 18),  2 July - 6 July 2018, Montpellier - FR.}

\author[label1]{R. L. Pearson}
\ead{r.l.pearson@sms.ed.ac.uk}
\address[label1]{The Higgs Centre for Theoretical Physics, University of Edinburgh, JCMB, KB, Mayfield Rd, Edinburgh EH9 3JZ, Scotland}

\author[label2]{C. Voisey\fnref{fn1}}
\fntext[fn1]{Speaker.}
\ead{voisey@hep.phy.cam.ac.uk}
\address[label2]{Cavendish Laboratory, University of Cambridge, Cambridge CB3 0HE, United Kingdom}

\begin{abstract}
An important limitation in current fits of parton distribution functions (PDFs) is that PDF uncertainties do not include any source of theoretical uncertainty. Here we present a general method for incorporating theoretical uncertainties into PDF fits, focussing in particular on perturbative missing higher order uncertainties (MHOUs). We consider two methods for estimating the effect of MHOUs on PDFs, both based on scale variations. Firstly, we present PDF fits based on theoretical predictions with varied scales, and use these to estimate the associated MHOUs. Secondly, we discuss the construction of a theoretical covariance matrix using scale variations, and its combination with the experimental covariance matrix currently used in PDF fits.
\end{abstract}


\end{frontmatter}

\section{Introduction} \label{sec:intro}

PDFs are crucial ingredients for making theoretical predictions for experiments such as the Large Hadron Collider (LHC). For the success of the LHC programme it is therefore important that PDFs are accurate and that their uncertainties are estimated in a reliable way. Here we address the current lack of a systematic inclusion of theoretical uncertainties in PDF fits; at present global PDF fits account only for experimental uncertainties, which propagate from the input experimental data, and methodological uncertainties, introduced by the choice of fitting procedure, e.g. see \cite{Ball:2017nwa}.

In the past, experimental uncertainties in PDF fits were large enough that the effects of many sources of theoretical uncertainties could be assumed to be negligible. However, we are now in an era of high-precision experimental measurement at the LHC \cite{Gao:2017yyd, Butterworth:2015oua, Rojo:2015acz}, and this is no longer the case. High-precision data have resulted in high-precision PDFs in kinematic regions that are spanned by these data. For example, in Ref.~\cite{Ball:2017nwa} (see Fig. 5.9) it is shown that parton luminosities constructed using NNPDF's latest global PDF set, NNPDF3.1, at next-to-next-to-leading order (NNLO) have a nominal precision that is as low as $1\%$ in some regions of phase space.

One type of theoretical uncertainty is MHOUs, which arise from the truncation of the perturbative expansions used in fixed order QCD calculations. In Ref.~\cite{Ball:2017nwa} (see Fig.~3.13), estimates of  MHOUs at NLO are obtained by computing the shift between the central values of the NLO and NNLO PDFs. These estimates of MHOUs are of comparable order to the PDF uncertainty at NLO, suggesting that including MHOUs in PDFs could now have a significant effect.

MHOUs are also of particularly increasing relevance because they are often the dominant source of theoretical uncertainty for LHC processes. An important example for which this is the case is Higgs production via gluon-gluon fusion \cite{deFlorian:2016spz}.

As a result of their importance, we focus here on a discussion of how to estimate the effect of MHOUs on PDFs. We emphasise, however, that the approach we ultimately present in Sect.~\ref{sec:thcovmat} is fully general and can be used to account for other types of theoretical uncertainty such as those due to nuclear effects.

\section{Scale variations} \label{sec:scalevariations}

Before moving on to our proposed frameworks for estimating MHOUs in PDF fits, we will review the standard technique for estimating MHOUs: scale variations. This method is based on the two unphysical scales that arise in fixed order QCD calculations: the renormalisation scale, $\mu_R$, and the factorisation scale, $\mu_F$. If all orders in the perturbative expansions are included, observables do not depend on either of these two scales. However, at fixed order there is a residual scale dependence. As a result, varying these scales in fixed order QCD calculations leads to different predictions, the spread of which quantifies the MHOU. In practice, calculations are done over a chosen range of scale combinations. For a discussion of the origin of these scales see Ref.~\cite{PDG_QCD}, and to see how fixed order QCD predictions depend on them see Refs.~\cite{forte_ridolfi, Altarelli:2008aj, vanNeerven:2000uj}, for example.

In order to estimate MHOUs for LHC processes, most studies adopt a prescription based on 7-point scale variations. For instance, this is the recommendation in the LHC Higgs Cross Section Working Group Yellow Report 4 for estimating the MHOU on a single theoretical prediction \cite{deFlorian:2016spz}. In this prescription, theoretical predictions are calculated for seven scale combinations, in which the baseline values for the two scales are varied by factors of 2 and $\frac{1}{2}$. Defining $\xi \equiv (\mu_R, \mu_F)$, these combinations are
\begin{equation}
\begin{split}
\bigg(\frac{\xi}{\xi_0}\bigg)_{\rm 7-point} = \{ (0.5, 0.5), (0.5, 1), (1, 0.5), (1, 1),\\(1, 2), (2, 1), (2, 2) \}\,,
\end{split}
\end{equation}
where $\xi_0$ represents the central (baseline) scales about which one is calculating variations. The combinations $\{(0.5, 2), (2, 0.5)\}$ are omitted, typically on grounds that doubling one scale while the other is halved corresponds to an unphysical separation between them. The envelope of the resulting seven predictions gives an estimate for the MHOU.

Scale variations is a well-established technique for estimating MHOUs for data points from a single physical process. However, extending this formalism to PDFs carries additional complications. Firstly, PDF fits use data from multiple physical processes, necessitating a framework that can consistently account for correlations between the theoretical predictions for different processes. Specifically, all theoretical predictions share common PDFs so MHOUs from the PDF evolution will be fully correlated, while there will also be additional correlations between theoretical predictions for similar processes, which have a renormalisation scale in common. For example, the theoretical predictions for jets from ATLAS will be correlated with those for jets from CMS due to both the PDF evolution and the shared renormalisation scale. Secondly, in global PDF fits we deal with $\sim 4000$ data points \cite{Ball:2017nwa}, so a computationally efficient procedure must be developed to compute theoretical predictions for several scale combinations for each data point.

In what follows we apply scale variations to PDFs predominantly by using and extending 7-point scale variations. We also consider 3-point scale variations in which, for one data point, $\mu_R$ and $\mu_F$ are set equal and varied in the same range as with 7-point scale variations. That is,
\begin{equation}
\begin{split}
\bigg(\frac{\xi}{\xi_0}\bigg)_{\rm 3-point} = \{ (0.5, 0.5), (1, 1), (2, 2) \}\,.
\end{split}
\end{equation}

\section{Fits with scale variations} \label{sec:scalevarfits}

We now turn our attention to using scale-varied theoretical predictions in PDF fits, and to estimating MHOUs at the PDF-level using the 3 and 7-point prescriptions discussed above. Computation of the necessary scale-varied deep-inelastic scattering (DIS) structure functions and hadronic cross sections is carried out using \texttt{APFEL} \cite{Bertone:2013vaa} and \texttt{APFELgrid} \cite{Bertone:2016lga}, which allow for arbitrary scale choices.

The fits we present use the NNPDF3.1 settings \cite{Ball:2017nwa} except for some minor changes to the selection of input data. Notably, we raise the minimum scale required for the input data from $Q^2=2.69$~GeV$^2$ to $Q^2=12.6$~GeV$^2$. This ensures that we preserve the perturbative convergence of predictions computed with $\mu_R$ and $\mu_F$ equal to half of their respective baseline values. Fits are produced for the case where both $\mu_R$ and $\mu_F$ are fully correlated across all data points.

\begin{figure}[!htb]
    \includegraphics[width=\linewidth]{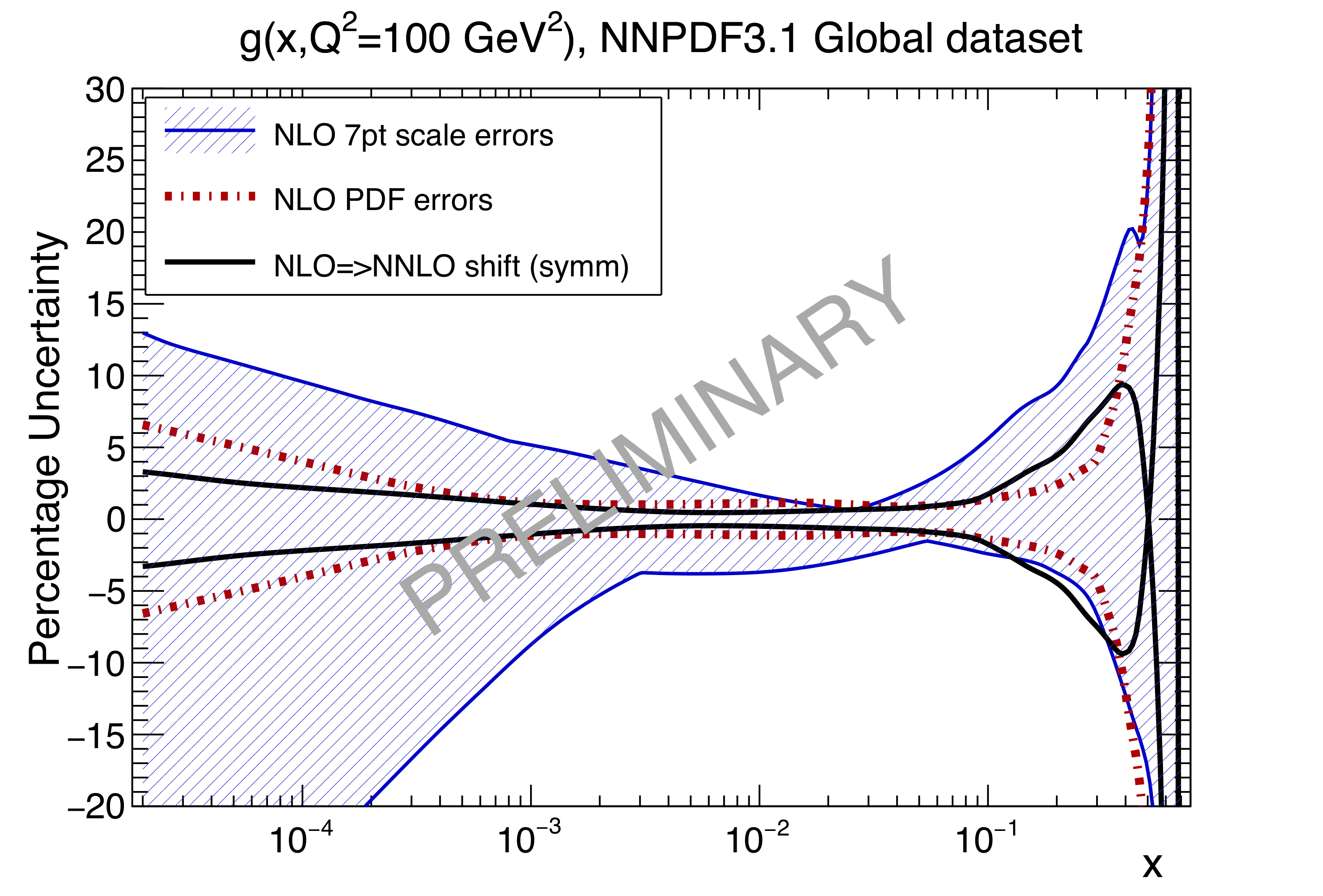}
    \includegraphics[width=\linewidth]{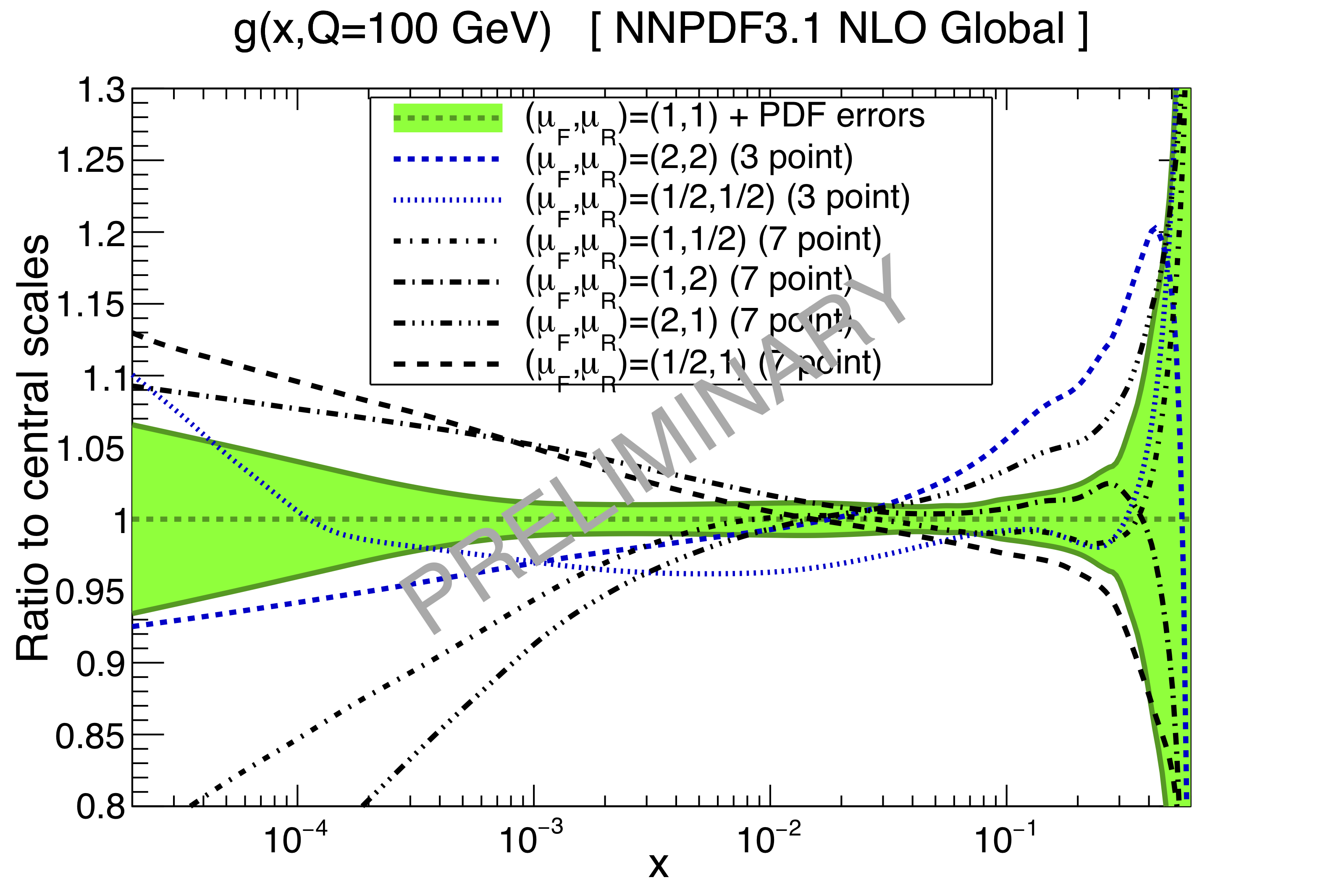}
    \caption{Preliminary results for 7-point scale variations for the gluon PDF, computed using a global data set at NLO and an energy scale of $Q^2=100$~GeV$^2$. The upper panel compares the PDF uncertainty (red), the symmetrised NLO-NNLO shift (black) and the scale uncertainty (blue). The scale uncertainty band is calculated by taking the envelope of the PDFs computed at each scale combination (for 7-point variations), as shown in the lower panel. The PDFs are normalised to the PDF computed at the central scales, which is shown with errors (green band). Scale choices corresponding to 3-point variations are coloured blue.}
\label{fig:scalevar_shift_comparisons}
\end{figure}

Fig.~\ref{fig:scalevar_shift_comparisons} shows some preliminary results from the fits with scale variations. In the upper plot we compare three uncertainty bands on the NLO gluon PDF. PDF errors (red dot-dashed) are plotted alongside two measures of MHOU: scale errors (blue hatched), determined using the procedure outlined above, and the (symmetrised) shift between NLO and NNLO predictions (black). This latter gives an indication of the ``true'' NLO uncertainty, and so we expect that a good estimation of MHOUs should be comparable to this shift, which provides a useful validation technique. We see that in general for 7-point variations the scale uncertainty encompasses the NLO-NNLO shift.

The lower plot of Fig.~\ref{fig:scalevar_shift_comparisons} provides insight into the origin of these scale errors. Here we compare the central values of the gluon PDF, calculated for each of the standard 7-point scale choices. Each PDF is normalised to the central scale result (green band). The scale combinations corresponding to 3-point variations are shown in blue. It is evident that the choice of scale can lead to large differences in PDF central values, especially at low $x$. The 3-point variations alone do not well represent the extent of these variations. In particular, we note that the scale uncertainty band can be tuned in three ways: by changing the prescription for scale variations (e.g. 3 versus 7-point); by changing the range in which the scales are varied; and by changing the way that the band is defined (i.e. by considering alternatives to taking the envelope).

An important point to note is that here both $\mu_R$ and $\mu_F$ are fully correlated between all data points.  This demonstrates a limitation of the envelope method for scale variations as presented above; as previously discussed, for a global data set correlations of renormalisation scales should be considered only between points belonging to the same process. To address this, separate fits could be performed using data from each process type, and a total envelope taken. However, this approach would necessitate producing increasingly many fits, which would become cumbersome. Furthermore, the envelope approach is not ideal since it lacks a solid statistical interpretation; for example, the envelope of two distinct uncertainties is not equivalent to the combination of the two uncertainties in a statistical sense.

\section{The theoretical covariance matrix} \label{sec:thcovmat}

To address these issues, we would like to incorporate theoretical uncertainties directly within PDF uncertainties at the fitting-level in an efficient and statistically sound way. To do this consistently we need to correctly account for theoretical correlations between the theoretical predictions for the different data points. We now outline a framework, which will be discussed in more detail in \cite{NNPDF_preparation}, through which arbitrary theoretical uncertainties and their correlations can be incorporated into future NNPDF uncertainty bands in a way that addresses the aforementioned issues. This is done via the construction of a theoretical covariance matrix in analogy with the experimental covariance matrix currently used in NNPDF fits. This will allow us to combine theoretical and experimental uncertainties when performing a fit.

In Ref.~\cite{Ball:2018odr} it is shown that by assuming that a particular source of theoretical uncertainty is Gaussian in nature and independent of the experimental uncertainty, this uncertainty can be in used in a $\chi^2$-minimisation in the same way as experimental uncertainties. That is, instead of minimising the usual `experimental'  $\chi^2$, which we write as
\begin{equation} \label{exp_chi2}
    \chi_{\rm exp}^2 = (y - t)^T \cov_{\rm exp}^{-1} (y - t)\,,
\end{equation}
we can minimise the following `total' $\chi^2$
\begin{equation} \label{total_chi2}
    \chi_{\rm tot}^2 = (y - t)^T (\cov_{\rm exp} + \cov_{\rm th})^{-1} (y - t)\,.
\end{equation}
In the above, $y$ is a vector of $N_{\rm data}$ data points to be fitted, and $t$ is a vector of theoretical predictions corresponding to $y$, which we calculate to some fixed order in perturbation theory. $\cov_{\rm exp}$ is the covariance matrix provided by experimentalists and $\cov_{\rm th}$ is a theoretical covariance matrix, to be determined. For a generic choice of scale combinations we propose to compute the theoretical covariance matrix according to the formula proposed in Ref.~\cite{Ball:2018odr}:
\begin{equation} \label{theoryunc1}
	\cov_{\text{th},\,ij} = \langle(t[\xi]-t[\xi_0])_i(t[\xi]-t[\xi_0])_j\rangle\,,
\end{equation}
where $i$ and $j$ index data points, $t$ is a theoretical prediction that depends on $\xi$, and the angled brackets denote the averaging over a given range of $\xi$ according to some user-defined prescription. The method of computing the average thus carries some ambiguity; for a generic multi-process case one needs to consider the different independent scale variations to be averaged over.

We now outline the extension of scale variations to multiple processes. Consider $n$ processes, each with an associated vector of theoretical predictions $t_p(\mu_0, \mu_p)$ (where $p$ = 1, ..., $n$, and in what follows we index two distinct processes with $p$ and $q$). Each process carries its own renormalisation scale, $\mu_p$, originating from the corresponding coefficient function or hard cross section. There is also a scale $\mu_0$, associated with the PDF evolution, which applies to all processes in the same way.

We can define the vector of differences between predictions calculated at two arbitrary scales and at the central scales by
\begin{equation} \label{eq:diff_vector}
  \Delta_p(\mu_0, \mu_p) \equiv t_p(\mu_0, \mu_p) - t_p(\mu_{0,\rm central},\mu_{p,\rm central})\,.
\end{equation}
Note that although we have $n$ processes in total, $\cov_{\text{th}}$ has two types of entry:
\begin{enumerate}
\item those which refer to points from the same physical process, and so have both correlated $\mu_0$ and $\mu_p$ variations, and
\item those which refer to points from two different physical processes, and so only have correlated $\mu_0$ variations.
\end{enumerate}
This means that we need only consider these two cases in order to construct the entire covariance matrix. Following this logic, we can define a sub-matrix by taking the outer product of the two difference vectors:
\begin{equation}
  \cov_{\text{th},\,pq}(\mu_0, \mu_p, \mu_q) \equiv N \sum_{\substack{\text{scale} \\ \text{choices}}}\Delta_p(\mu_0, \mu_p)\Delta_q(\mu_0, \mu_q)\,,
\end{equation}
where $N$ is an overall normalisation factor determined by averaging over variations for each scale and then adding the contributions from all the independent scales. Note that here the combinations of $\Delta$s are implicitly outer products.

\begin{figure*}[!htb]
    \centering
    \includegraphics[width=\linewidth]{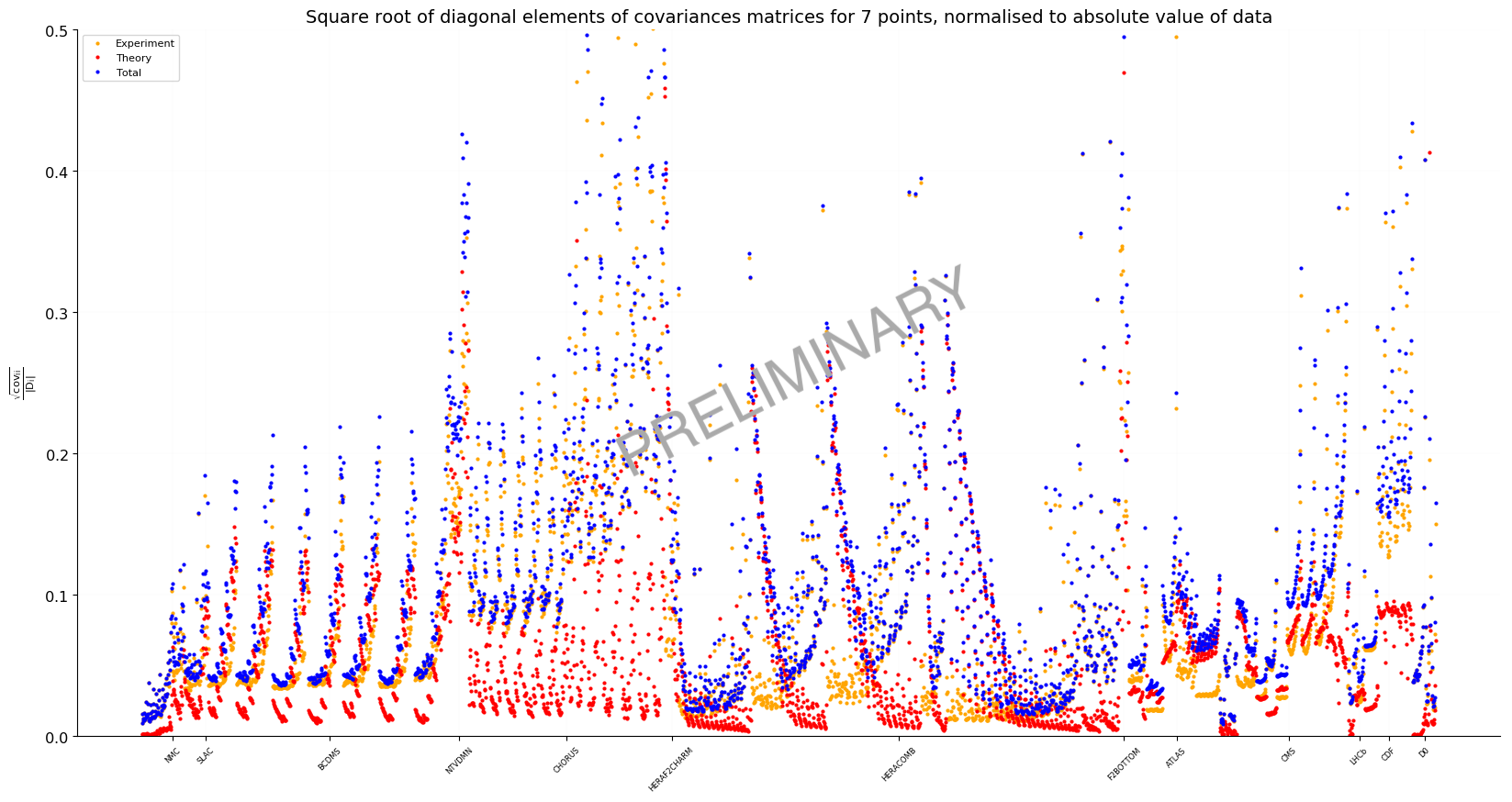}
    \caption{A plot of the relative uncertainties associated with each data point for a global data set, grouped according to experiment and, within this, according to kinematic region. The uncertainties are evaluated as the square root of the diagonal elements of the relevant covariance matrix, normalised to the absolute values of the data. Yellow points correspond to experimental uncertainties, red points to MHOUs at NLO determined from 7-point scale variations, and blue points to the combined uncertainty. Regions with red points close to the $x$-axis thus correspond to areas where experimental uncertainties dominate, and converseley regions with yellow points close to the $x$-axis correspond to domination of theoretical errors.}
\label{fig:diag}
\end{figure*}

As an initial assessment of MHOUs estimated using this method it is instructive to first look at the errors associated with each individual data point. To do this we consider the diagonal elements of the covariance matrices, as in Fig.~\ref{fig:diag}. Here we display the square roots of the diagonal elements normalised to data, which is a measure of the relative error for that point. Yellow points correspond to experimental uncertainties, red points to 7-point MHOUs, and blue points to the combined uncertainty. For many of the older experiments such as BCDMS and CHORUS we see that experimental uncertainties are highly dominant. However, for other data like the HERA combined structure function data (represented as `HERACOMB' in the plots), and the ATLAS and CMS data, MHOU uncertainties are larger than experimental uncertainties in some regions. This is in line with what we might expect. Note that these MHOUs are for NLO and will in general be smaller at NNLO.

\begin{figure*}[!htb]
    \centering
    \includegraphics[width=0.49\linewidth]{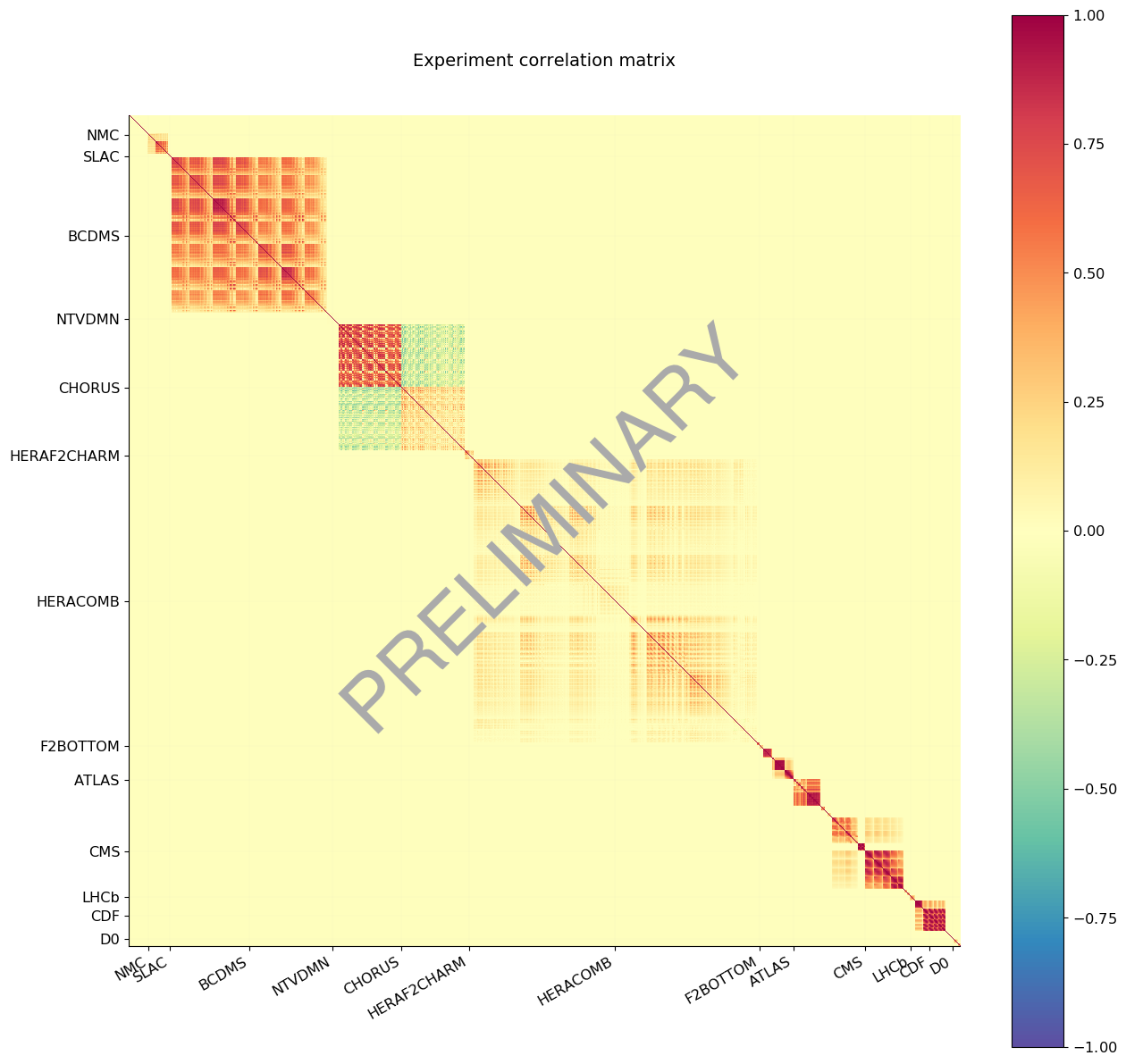}
    \includegraphics[width=0.49\linewidth]{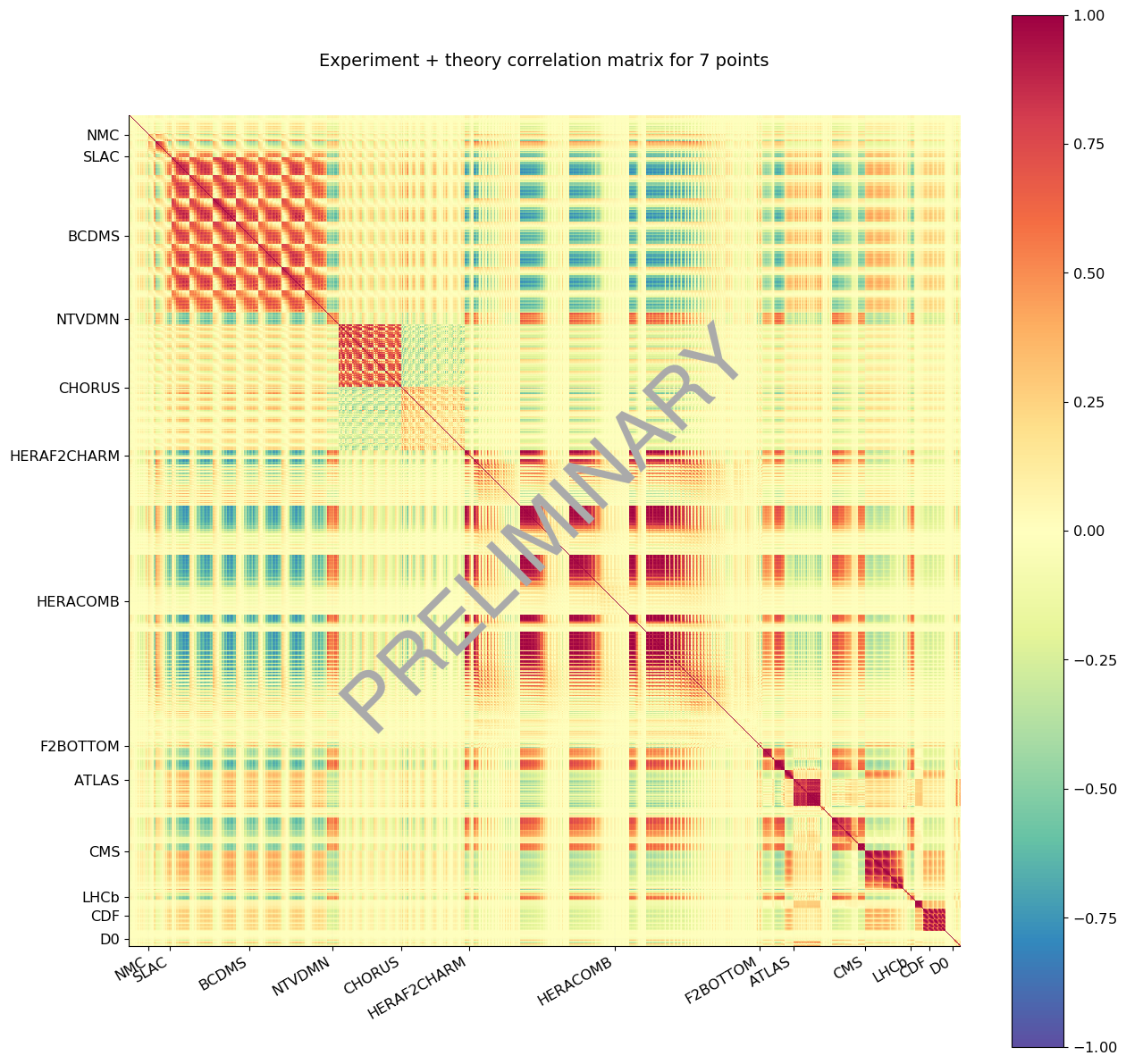}
    \caption{A graphical representation of the experimental correlation matrix (left), and the total correlation matrix (right). The total correlation matrix is constructed by combining the theoretical and experimental covariance matrices and then applying Eq.~\ref{eqn:corr}. Results are for a global data set at NLO, with the theoretical covariance matrix constructed using multi-process 7-point scale variations. Processes are categorised as: DIS, Drell-Yan, jets and heavy quarks. The $x$ and $y$-axes both run over all $N_{\rm data}$ data points, which are grouped by experiment, and within experiments according to kinematics. The entries range between $-$1 (dark blue, maximally anticorrelated) and $+$1 (dark red, maximally correlated).}
\label{fig:corrmats}
\end{figure*}

We can define correlation matrices by
\begin{equation} \label{eqn:corr}
  \corr_{ij} = \frac{\cov_{ij}}{\sqrt{\cov_{ii}} \sqrt{\cov_{jj}}}\,,
\end{equation}
which allow us to analyse the correlation structure between data points more easily than by using covariance matrices. The experimental correlation matrix and the total correlation matrix, which is calculated by following the prescription described above and adding together the experimental and theoretical covariance matrices, are displayed in Fig.~\ref{fig:corrmats}. The results presented here are based on a scheme of four different process types: DIS, Drell-Yan, jets and heavy quarks. Notice that $\corr_{\text{exp}}$ is block-diagonal in form, as there are no experimental correlations between data points from different experiments. Darker regions correspond to a dominance of systematic errors over statistical errors. We can see that for example BCDMS is systematics-dominated whereas HERACOMB is statistics-dominated. For some LHC processes the data are already dominated by systematic errors.

The inclusion of MHOUs introduces off-diagonal entries due to shared factorisation and renormalisation scale variations. The effect on the correlation matrix provides visual confirmation of the significance of theoretical errors when compared with experimental errors. We can also investigate correlations between different data sets. Structure along the block-diagonal reflects correlations within each experiment, while entries elsewhere quantify the correlation between scale variation errors in different experiments. The strongest reinforcement along the block-diagonal is seen for the HERACOMB data. This suggests that scale variations between data points within HERACOMB are more highly correlated than within other experiments. Off the block-diagonal we see a rich correlation structure. Taking the interaction of BCDMS with other data sets as an example, we see that the data from BCDMS and HERACOMB are largely anticorrelated, the data from BCDMS and CHORUS are largely correlated, while the data from BCDMS and the hadronic experiments has a more mixed correlation structure. We can interpret this in terms of the kinematic range of the different data sets; those in a similar kinematic range will show strong correlations, and those in different ranges will show anticorrelations. For example, for the HERACOMB and BCDMS data, which are from different $x$ regions, this has its origin in the sum rules; if the PDF decreases at large $x$ it must increase at small $x$ to compensate.

Finally, we can use Eq.~\ref{total_chi2} to assess the impact of MHOUs on the value of the $\chi^2/N_{\rm data}$. We find that this reduces from 1.123, with $\cov_{\text{exp}}$ only, to 1.111, with $\cov_{\text{exp+th}}$. Such a reduction is expected, as we have introduced additional errors. If we exclude all theoretical correlations between different data points, i.e. only consider the diagonal elements of $\cov_{\text{th}}$, we obtain 0.579. This tells us that, when included, the theoretical correlations heavily compensate for the addition of an extra source of uncertainty, i.e. the MHOUs, at the $\chi^2$-level. This highlights the importance of correlations in addition to the size of the MHOUs. We note that the true test of the impact of MHOUs at the $\chi^2$-level will be seen when recalculating these values using the PDFs that are fitted using $\cov_{\text{exp+th}}$.

\section{Conclusions and future work} \label{sec:conc}

We have shown that we can estimate MHOUs in PDFs by producing PDF fits with scale-varied theoretical predictions. We have also presented a framework for including MHOUs in future NNPDF fits, in which MHOUs are propagated to PDF uncertainties by defining a theoretical covariance matrix. The next step in this work is to produce such PDF fits. These will be the first fits of their kind, and will represent an important improvement in PDF determinations. Using these fits the impact of different choices for calculating scale variations will be further investigated; in particular, we will investigate how the choices of prescription, the range of scale variation, and the categorisation of process types affect PDF fits.

Finally, this work is clearly extendable to other sources of theoretical uncertainty, via the definition of additional corresponding theoretical covariance matrices, as well as being extendable for use in other NNPDF fits, such as fits of fragmentation functions or polarised PDFs.\\

\textbf{Acknowledgements}

We would like to thank the QCD 18 conference organisers. We would also like to thank our collaborators in the NNPDF Collaboration for their involvement in this work. This work was funded by the Science and Technology Facilities Council grants ST/R504737/1, ST/N50399X/1 and ST/R504671/1.

\end{document}